\documentclass[12pt]{article}
\usepackage{amsfonts,graphicx,amssymb,amsmath,amsthm,epsfig}
\usepackage{float}
\allowdisplaybreaks
\begin{document}
\title{\textbf{Analysis of Charged Compact Stars with Bardeen Black Hole in
$f(\mathfrak{Q}, \mathcal{T})$ Gravity}}
\author{M. Sharif \thanks {msharif.math@pu.edu.pk}~and Iqra Ibrar
\thanks{iqraibrar26@gmail.com}\\
Department of Mathematics and Statistics, The University of Lahore,\\
1-KM Defence Road Lahore-54000, Pakistan.}
\date{}
\maketitle
\begin{abstract}
This study investigates the behavior of charged compact stars within
the $f(\mathfrak{Q}, \mathcal{T})$ gravitational framework, where
$\mathfrak{Q}$ denotes the non-metricity scalar and $\mathcal{T}$
represents the trace of the energy-momentum tensor. Recognized as a
promising model for explaining the accelerated expansion of the
universe, this approach offers a strong theoretical foundation. A
central focus of this research is the application of Bardeen's model
to describe the exterior spacetime. Additionally, the study examines
the internal structure of compact stars using a solution based on
the Finch-Skea metric potential. Various physical properties,
including energy density, pressure components, anisotropy, energy
conditions and equation of state parameters are analyzed through
graphical representations. Equilibrium conditions are explored via
the Tolman-Oppenheimer-Volkoff equation while key characteristics
such as the mass-radius relationship, compactness, surface redshift
and stability criteria based on the causality condition and
adiabatic index are thoroughly evaluated. The analysis concludes
that the proposed solutions for charged compact stars in this
framework are both theoretically consistent and physically valid.
\end{abstract}

\section{Introduction}

The late-stage evolution of celestial formation driven by intense
gravitational forces has been a focal point in astrophysical
research, advancing our understanding of these phenomena. Baade and
Zwicky \cite{a} first proposed that supernovae could produce dense
stellar remnants, paving the way for the study of compact stars.
This theory gained support in 1967 when Bell and Hewish discovered
pulsars, highly magnetized, rapidly rotating neutron stars, marking
a pivotal shift in understanding star transformations into compact
objects. These discoveries have significantly broadened and refined
the classification of stars. Compact objects form when nuclear
fusion ceases, depleting fuel and allowing gravitational collapse.
White dwarfs achieve stability through electron degeneracy pressure,
while neutron stars rely on neutron degeneracy pressure. Black
holes, however, collapse entirely, resulting in singularities and
event horizons. These stellar remnants are marked by their extreme
density and small radii, a critical phase in stellar evolution
shaped by internal pressures and gravitational forces.

Compact stars, with immense mass and minuscule radii, continue to
fascinate researchers. Despite progress, many questions remain
unanswered. Early models assumed isotropic fluid behavior. Ruderman
\cite{b} introduced internal anisotropy, bringing new insights into
stellar structures. This led to studies using the equation of state
(EoS) parameter under anisotropic pressure. Mak and Harko \cite{c}
investigated charged strange stars through conformal motion. Rahaman
et al. \cite{d} analyzed the Krori-Barua metric for modeling strange
stars, exploring its mathematical implications. Singh and Pant
\cite{e} proposed solutions for anisotropic stellar objects within
energy conditions and the Buchdahl limit. Maurya et al. \cite{f}
developed Einstein-Maxwell solutions for spherically symmetric
objects, modeling the strange star Her X-1 and revealing stability
mechanisms involving electromagnetic and anisotropic forces.

Building on compact stellar studies, Bardeen black hole offers a
singularity-free model of regular black holes. Proposed by James
Bardeen \cite{g} in 1968, it incorporates a non-linear
electromagnetic field that modifies spacetime geometry. As a
solution to Einstein field equations, this model features a
singularity-free core. Studies of its stability, causal structure
and quantum gravity implications bridge general relativity (GR) and
modern quantum theories. Moreno and Sarbach \cite{h} validated
stability conditions for regular black hole models. Ulhoa \cite{i}
analyzed gravitational perturbations and quasinormal frequencies of
the Bardeen black hole. Fernando \cite{j} studied the Bardeen-de
Sitter black hole's greybody factors and absorption cross sections
for scalar fields based on its mass, charge and cosmological
constant.

Modified gravitational theories (MGTs) have been developed to
explain cosmic phenomena such as dark matter and the accelerated
expansion of the universe \cite{k}. These theories aim to address
the shortcomings of GR by incorporating additional components, like
dark energy (DE), to reconcile observable discrepancies. By
expanding GR through elements such as the Ricci scalar curvature,
MGTs tackle issues related to the universes expansion and dynamics,
emphasizing the importance of alternative approaches for a deeper
understanding of cosmology. Various modifications to GR have been
introduced in the scientific literature to overcome its limitations.
One notable advancement is $f(\mathfrak{Q})$ gravity, which employs
the non-metricity scalar as a central factor in gravitational
interactions \cite{n}. This framework is grounded in Weyl geometry,
an extension of Riemannian geometry that forms the basis of GR.
Unlike Riemannian geometry, which associates gravitational effects
with directional changes during vector parallel transport, Weyl
geometry attributes these effects to changes in vector length. This
novel approach offers profound insights into the fundamental
mechanisms of gravitational interactions.

Lazkoz et al. \cite{3a} reformulated $f(\mathfrak{Q})$ gravity using
redshift as a parameter to assess its feasibility as an alternative
explanation for late-time cosmic acceleration. Soudi \cite{3b}
examined the polarization of gravitational waves within this
framework, while additional theoretical and observational studies
were detailed in \cite{3c}. Bajardi et al. \cite{3d} employed the
order reduction method to constrain models and derive bouncing
cosmologies. Shekh \cite{3e} investigated the dynamics of
holographic DE models using this theory. Frusciante \cite{3f}
proposed a specific model under $f(\mathfrak{Q})$ gravity,
highlighting foundational parallels with the $\Lambda$CDM model. Lin
and Zhai \cite{3g} explored its impact on spherically symmetric
systems, deriving solutions for compact stars. Atayde and Frusciante
\cite{3h} applied observational data from the cosmic microwave
background and type-Ia supernovae to place constraints on this
theory. Lymperis \cite{3i} studied cosmic evolution with and without
a cosmological constant in the presence of phantom DE, a topic also
addressed by Dimakis et al. \cite{3j}. Khyllep et al. \cite{3k}
analyzed $f(\mathfrak{Q})$ gravity, demonstrating their potential as
strong alternatives to the $\Lambda$CDM framework. Recent work
\cite{3l} has delved deeply into the geometrical and physical
implications of this gravity, offering significant insights into its
foundational aspects.

Xu et al. \cite{6-a} expanded the $f(\mathfrak{Q})$ framework by
integrating the trace of the energy-momentum tensor (EMT) into the
functional action, leading to $f(\mathfrak{Q}, \mathcal{T})$
gravity. This modification creates a direct relationship between
matter distribution and spacetime geometry, introducing additional
degrees of freedom. The theory provides a plausible explanation for
cosmic acceleration without invoking DE, unifying early-universe
inflation and late-time acceleration within a single cohesive
framework. By redefining the geometry of spacetime, it broadens the
scope of GR. Unlike alternative models that often depend on
fine-tuning or auxiliary fields, $f(\mathfrak{Q}, \mathcal{T})$
enhances the adaptability of cosmological theories, demonstrating
consistency with observational data and shedding light on
large-scale cosmic dynamics. Its second-order field equations
simplify calculations, making it a practical and efficient tool for
cosmological research. Although this approach offers valuable
insights into the universe evolution, further studies are needed to
confirm its potential as a comprehensive cosmological framework.

Recent studies have delved deeply into the cosmological significance
of this theoretical framework, highlighting its potential to address
pressing questions about DE and the universe accelerated expansion.
Shiravand et al. \cite{13g} explored its application to cosmic
inflation. Pati et al. \cite{3-c} focused on the time evolution of
key cosmological parameters. Narawade et al. \cite{3-f} investigated
the dynamics of accelerating cosmological models, affirming their
stability and viability. Gadbail et al. \cite{1a} analyzed the
evolutionary patterns of the universe under this modified paradigm,
shedding light on phenomena like accelerated expansion. Bourakadi et
al. \cite{1c} studied the model role in the formation of primordial
black holes, establishing a link with inflationary processes. Shekh
\cite{1d} introduced an innovative scale factor for modeling
late-time acceleration and used statefinder diagnostics to evaluate
its performance. Venkatesha et al. \cite{3-e} examined the
possibility of traversable wormholes emerging from this extended
gravity theory. Gul et al. \cite{25} investigated the structural and
physical properties of compact stellar objects within this context.
Khurana et al. \cite{26} proposed a time-dependent, higher-order
formulation for the deceleration parameter, refining its accuracy
with observational data. Furthermore, recent investigations
\cite{14a} have extensively explored $f(\mathfrak{Q}, \mathcal{T})$
gravity, revealing its diverse geometrical and physical
implications.

The Finch-Skea metric was developed to overcome the challenges
associated with modeling compact astrophysical objects \cite{37}.
Finch and Skea \cite{38} enhanced this model to provide a more
precise representation of compact stellar objects, highlighting its
non-singular nature at the stellar core and its compliance with
essential physical conditions, including energy conservation and
causality principles. Bhar \cite{45} utilized this metric together
with the Chaplygin gas EoS to analyze the distinctive features of
strange stars. This metric is particularly significant for modeling
the structure of compact stars in spacetimes with four or more
dimensions \cite{44}. It has also been widely applied to study
gravastars, wormholes and strange stars in GR and MGTs \cite{gh}.
Malik et al. \cite{4a7a} investigated charged compact stars within
$f(\mathfrak{R}, \mathcal{T})$ gravity theory using Bardeen model
and Finch-Skea metric to analyze their structure, physical
parameters and stability properties, where $\mathfrak{R}$ is a Ricci
scalar. Sharif and Naz \cite{1gg} investigated gravastars using this
metric under the framework of $f(\mathfrak{R}, \mathcal{T}^2)$
gravity. Sharif and Manzoor \cite{1hh} analyzed the equilibrium of
dense anisotropic stellar structures within the same framework.
Mustafa et al. \cite{3gg} studied the dynamics of anisotropic
compact stellar configurations with spherical symmetry in
$f(\mathfrak{Q})$ gravity, incorporating the Karmarkar condition
alongside the Finch-Skea framework. Karmakar et al. \cite{2gg}
presented a 5D model in MGT utilizing this ansatz, applied it to the
star EXO 1785-248 and demonstrated its compatibility with all
physical criteria. Interest in the Finch-Skea model in MGTs is
growing, as highlighted by recent studies \cite{4a9}.

To our knowledge, no previous research has been investigated for
charged compact structures employing the Finch-Skea metric alongside
the fascinating characteristics of Bardeen black hole within
$f(\mathfrak{Q}, \mathcal{T})$ gravity. This approach adds depth to
the analysis due to the unique nature of the Bardeen solution. The
paper is organized as follows. Section \textbf{2} outlines the
formulation of $f(\mathfrak{Q}, \mathcal{T})$ gravity in conjunction
with Maxwell equations. Using the Finch-Skea metric, the field
equations are derived for a spherically symmetric framework. Section
\textbf{3} evaluates the constants by applying the matching
conditions for connecting the interior and exterior spacetimes. In
section \textbf{4}, we provide an analysis of the physical
properties of the proposed model, along with graphical behavior.
Finally, section \textbf{5} provides a summary of our results.

\section{The $f(\mathfrak{Q},\mathcal{T})$ Formalism}

The action in the $f(\mathfrak{Q},\mathcal{T})$ framework, modified
to include the matter Lagrangian $\mathbb{L}_{m}$ and the
electromagnetic field Lagrangian $\mathbb{L}_{\mathbf{e}}$, is given
by the following expression \cite{6-a}
\begin{equation}\label{22}
S=\frac{1}{2\kappa}\int f(\mathfrak{Q},\mathcal{T})\sqrt{-g}d^{4}x+\int
(\mathbb{L}_{m}+\mathbb{L}_{\mathbf{e}})\sqrt{-g}d^{4}x.
\end{equation}
In this case, $\kappa$ represents the coupling constant while $g$
stands for the determinant of the metric tensor. The Lagrangian for
the electromagnetic field is formulated as
\begin{equation}
\mathbb{L}_{e} =\frac{-1}{16\pi} F^{\varsigma\eta}F_{\varsigma\eta}.
\end{equation}
The Maxwell field tensor can be written as $F_{\varsigma\eta} =
\psi_{\eta,\varsigma}-\psi_{\varsigma,\eta}$ with $\psi_{\varsigma}$
denoting the four potential. Moreover, $\mathfrak{Q}$ is
characterized as
\begin{equation}\label{23}
\mathfrak{Q}=-g^{\xi\vartheta}(\mathrm{L}^{\gamma}_{\varpi\xi}\mathrm{L}^{\varpi}_{\vartheta\gamma}
-\mathrm{L}^{\gamma}_{\varpi\gamma}\mathrm{L}^{\varpi}_{\xi\vartheta}),
\end{equation}
where
\begin{equation}\label{24}
\mathrm{L}^{\iota}_{\mu\varpi}=-\frac{1}{2}g^{\iota\lambda}
(\nabla_{\varpi}g_{\mu\lambda}+\nabla_{\mu}g_{\lambda\varpi}
-\nabla_{\lambda}g_{\mu\varpi}).
\end{equation}
The superpotential linked to $\mathfrak{Q}$ is represented as
\begin{align}\label{26}
P^{\iota}_{\gamma\eta}&=-\frac{1}{2}\mathrm{L}^{\iota}_{\gamma\eta}
+\frac{1}{4}(\mathfrak{Q}^{\iota}-\tilde{\mathfrak{Q}}^{\iota}) g_{\gamma\eta}-
\frac{1}{4} \delta ^{\iota}\;_{({\gamma}}\mathfrak{Q}_{\eta)},
\end{align}
where
\begin{align}\label{25}
\mathfrak{Q}_{\iota}= \mathfrak{Q}^{~\xi}_{\iota~\xi}, \quad
\tilde{\mathfrak{Q}}_{\iota}= \mathfrak{Q}^{\xi}_{\iota\xi}.
\end{align}
Furthermore, the formulation of $\mathfrak{Q}$ as derived from the
superpotential, is given by \cite{4-G}
\begin{align}\label{27}
\mathfrak{Q}=-\mathfrak{Q}_{\iota\gamma\eta}P^{\iota\gamma\eta}
=-\frac{1}{4}(-\mathfrak{Q}^{\iota\gamma\eta}\mathfrak{Q}_{\iota\eta\gamma}
+2\mathfrak{Q}^{\iota\eta\gamma}\mathfrak{Q}_{\gamma\iota\eta}
-2\mathfrak{Q}^{\eta}\tilde{\mathfrak{Q}}_{\eta}
+\mathfrak{Q}^{\eta}\mathfrak{Q}_{\eta}).
\end{align}

The field equations are found by evaluating the change in the action
with respect to the metric tensor and equating the resulting
expression to zero
\begin{align}\nonumber
\delta S&=0=\int \frac{1}{2\kappa}\delta
[\sqrt{-g}f(\mathfrak{Q},\mathcal{T})]+\delta[\sqrt{-g}\mathbb{L}_{m}]d^{4}x,
\\\nonumber
0&=\int \frac{1}{2\kappa}\bigg( \frac{-1}{2} f g_{\gamma\eta}
\sqrt{-g} \delta g^{\gamma\eta} + f_{\mathfrak{Q}} \sqrt{-g} \delta
\mathfrak{Q} + f_{\mathcal{T}} \sqrt{-g} \delta
\mathcal{T}\\\label{28} &-\frac{1}{2\kappa} \mathcal{T}_{\gamma\eta}
\sqrt{-g} \delta g^{\gamma\eta}\bigg)d^ {4}x.
\end{align}
We also define
\begin{align}\label{29}
\mathcal{T}_{\gamma\eta} &= \frac{-2}{\sqrt{-g}} \frac{\delta (\sqrt{-g}
\mathbb{L}_{m})}{\delta g^{\gamma\eta}}, \quad \Theta_{\gamma\eta}= g^{\iota\mu}
\frac{\delta \mathcal{T}_{\iota\mu}}{\delta g^{\gamma\eta}}.
\end{align}
As a result, we have $ \delta \mathcal{T} = \delta (\mathcal{T}_{\gamma\eta}
g^{\gamma\eta}) = (\mathcal{T}_{\gamma\eta} + \Theta_{\gamma\eta}) \delta
g^{\gamma\eta}$. Integrating the specified factors into Eq.\eqref{28} yields the
resulting outcome
\begin{eqnarray}\nonumber
\delta S =0&=&\int \frac{1}{2\kappa}\bigg\{\frac{-1}{2}f g_{\gamma\eta}\sqrt{-g}
\delta g^{\gamma\eta} + f_{\mathcal{T}}(\mathcal{T}_{\gamma\eta}+
\Theta_{\gamma\eta})\sqrt{-g} \delta g^{\gamma\eta}
\\\nonumber
&-&f_{\mathfrak{Q}} \sqrt{-g} (P_{\gamma\iota\mu} \mathfrak{Q}_{\eta}~^{\iota\mu}-
2\mathfrak{Q}^{\iota\eta}~_{\gamma} P_{\iota\mu\eta}) \delta
g^{\gamma\eta}+2f_{\mathfrak{Q}} \sqrt{-g} P_{\iota\gamma\eta} \nabla^{\iota} \delta
g^{\gamma\eta}
\\\label{30}
&+&2f_{\mathfrak{Q}}\sqrt{-g}P_{\iota\gamma\eta} \nabla^{\iota} \delta g^{\gamma\eta}
\bigg\}- \frac{1}{2\kappa} \mathcal{T}_{\gamma\eta}\sqrt{-g} \delta g^{\gamma\eta}d^
{4}x.
\end{eqnarray}
By integrating the term $2f_{\mathfrak{Q}} \sqrt{-g} P_{\iota\gamma\eta}
\nabla^{\iota} \delta g^{\gamma\eta}$ and imposing the relevant boundary conditions,
we derive $-2 \nabla^{\iota} (f_{\mathfrak{Q}} \sqrt{-g} P_{\iota\gamma\eta}) \delta
g^{\gamma\eta}$. Here, $f_{\mathfrak{Q}}$ and $f_{\mathcal{T}}$ represent the partial
derivatives of the function with respect to $\mathfrak{Q}$ and $\mathcal{T}$,
respectively. Consequently, the field equations are formulated as follows
\begin{eqnarray}\nonumber
\kappa(\mathcal{T}_{\gamma\eta}+\mathbb{E}_{\gamma\eta})&=&\frac{-2}{\sqrt{-g}}
\nabla_{\iota} (f_{\mathfrak{Q}}\sqrt{-g} P^{\iota}_{\gamma\eta})- \frac{1}{2} f
g_{\gamma\eta} + f_{\mathcal{T}} (\mathcal{T}_{\gamma\eta} +
\Theta_{\gamma\eta})\\\label{31} &-&f_{\mathfrak{Q}} (P_{\gamma\iota\mu}
\mathfrak{Q}_{\eta}~^{\iota\mu} -2\mathfrak{Q}^{\iota\mu}~_{\gamma}
P_{\iota\mu\eta}).
\end{eqnarray}

For the electromagnetic field, the stress-energy tensor is expressed
as
\begin{equation}
\mathbb{E}_{\gamma\eta} = \frac{1}{4\pi} \big( F^{\nu}_{\gamma} F_{\nu\eta} -
\frac{1}{4} g_{\gamma\eta} F_{\varsigma\nu} F^{\varsigma\nu} \big).
\end{equation}
Furthermore, the Maxwell equations are
\begin{equation}
(\sqrt{-g} F_{\gamma\eta})_{;\eta} = 4\pi J_\gamma \sqrt{-g} F_{[\gamma\eta;\delta]}
= 0.
\end{equation}
Here, $J_\gamma = \sigma u_\gamma$ represents the electric
four-current, with $\sigma$ denoting the charge density. The
electric field strength is described as
\begin{equation}
\mathbb{E}(r) = \frac{e^{\frac{\alpha + \beta}{2}}}{r^2}q(r).
\end{equation}
The function $q(r)$ represents the aggregate electric charge
contained inside a sphere with radius $r$ and is defined by
\begin{equation}
q(r) = 4\pi \int_0^r \sigma r^2 e^\beta dr.
\end{equation}
This leads to the expression for the charge density
\begin{equation}\label{20}
\sigma = \frac{e^{-\beta}}{4\pi r^2} \frac{dq(r)}{dr}.
\end{equation}

We will now analyze the interior metric characterized by spherical
symmetry. When formulated in spherical coordinates, the metric
follows the standard structure
\begin{equation}\label{1a}
ds^{2}=-e^{\alpha(r)}dt^{2}+e^{\beta(r)}dr^{2}+r^{2}(d\theta^{2}+\sin^{2}\theta
d\phi^{2}).
\end{equation}
To characterize the fluid distribution, we employ an anisotropic EMT
defined as
\begin{equation}\label{B}
\mathcal{T}_{ab} = (\rho + p_t) u_a u_b + p_t g_{ab} + (p_r - p_t) v_a v_b.
\end{equation}
Within this framework, the four-velocity of the fluid is represented
by $u_{a}$ and $v_a$ denotes the four-vector. The quantities $\rho$,
$p_{t}$ and $p_{r}$ denote the energy density, tangential pressure
and radial pressure, respectively. In the setting of
$f(\mathfrak{Q}, \mathcal{T})$ gravity, the Einstein-Maxwell field
equations are formulated as
\begin{align}\nonumber
\kappa\rho &=-f_{\mathfrak{Q}} \bigg(\mathfrak{Q}(r)+\frac{1}{r^2}+\frac{
\big(\alpha^\prime(r)+\beta^ \prime(r)e^{-\beta (r)}\big)}{r}\bigg)+\frac{2
f_{\mathfrak{Q}\mathfrak{Q}}  \mathfrak{Q}^\prime(r)e^{-\beta (r)}}{r}
\\\label{1c}
&-\frac{1}{3} f_{\mathcal{T}} (p_{r}+2 p_{t}+3 \varrho)+\frac{f}{2}-\frac{q^2}{r^4},
\\\label{1d}
\kappa p_{r}&=f_{\mathfrak{Q}}
\big(\frac{1}{r^2}+\mathfrak{Q}(r)\big)-\frac{f}{2}+\frac{2}{3} f_{\mathrm{T}}
(p_{t}-p_{r})+\frac{q^2}{r^4},
\\\nonumber
\kappa p_{t}&=f_{\mathfrak{Q}} \bigg[\frac{\mathfrak{Q}(r)}{2}- \big(\frac{\alpha^
{\prime\prime}(r)}{2}+\big(\frac{1}{2 r}+\frac{\alpha^\prime(r)}{4}\big)
\big(\alpha^\prime(r)-\beta^ \prime(r)\big)\big)e^{-\beta (r)}\bigg]
\\\label{1e}
&-f_{\mathfrak{Q}\mathfrak{Q}} e^{-\beta (r)} \mathfrak{Q}^\prime(r)
\bigg(\frac{\alpha^\prime(r)}{2}+\frac{1}{r}\bigg)+\frac{1}{3} f_{\mathcal{T}}
(p_{r}-p_{t})-\frac{f}{2}-\frac{q^2}{r^4}.
\end{align}
Furthermore, $\mathfrak{Q}$ can also be expressed as \cite{11a}
\begin{equation}\label{1b}
\mathfrak{Q}(r)=\frac{2e^{-\beta(r)}}{r}\bigg(\frac{1}{r}+\alpha^ \prime(r)\bigg),
\end{equation}
where prime is the derivative with respect to $r$. We utilize a
specific model of $f(\mathfrak{Q}, \mathrm{T})$ described as
\cite{10-b}
\begin{equation}\label{2a}
f(\mathfrak{Q},\mathcal{T})=\zeta  \mathfrak{Q}+\eta  \mathcal{T}.
\end{equation}

With $\zeta$ and $\eta$ as non-zero arbitrary constants, this
cosmological model has been extensively explored in studies
\cite{10-c}. Its selection is driven by its simplicity,
computational efficiency and capacity to describe the fundamental
physics of compact stars. The parameters $\zeta$ and $\eta$ govern
the contributions from non-metricity and the trace of the EMT,
respectively, under the linear constraints $f_\mathfrak{Q} = \zeta,
f_{\mathfrak{QQ}} = 0, f_\mathcal{T} = \eta$ and $f_{\mathcal{TT}} =
0$. This linear structure simplifies the field equations, enabling
analytical solutions while preserving their physical significance.
The model captures matter geometry interactions and provides
reliable predictions for critical astrophysical characteristics,
such as the structure and dynamics of compact stars. Its simplicity
and practicality make it widely recognized in the literature as a
robust framework for further exploration.

We employ the Finch-Skea solution for its clear and efficient
approach in developing accurate models of the interior spacetime.
This choice is motivated by its stable characteristics and its
ability to meet fundamental standards of acceptability \cite{x1}.
Over time, numerous researchers have extensively expanded the
Finch-Skea isotropic model to investigate various astronomical
phenomena. Different gravitational theories have utilized the
Finch-Skea ansatz to model various astrophysical systems. In this
study, the Finch-Skea ansatz and its corresponding solution play a
pivotal role in our methodology. Its regular behavior at the core
and smooth transition from the interior to the exterior regions make
it well-suited for analyzing dense celestial bodies. The Finch-Skea
solutions are mathematically represented using the Adler-Finch-Skea
metric potentials \cite{x2} as follows
\begin{equation}\label{3a}
\alpha(r)=\log[\chi(1+\varphi r^2)^2],\quad \beta(r)=\log[1+16\chi\varphi^2\gamma
r^2].
\end{equation}
In this context, the constants $\chi$, $\varphi$ and $\gamma$ are
undetermined and can be resolved through matching conditions. By
applying Eqs.\eqref{2a} and \eqref{3a} into
Eqs.\eqref{1c}-\eqref{1e}, we obtain the corresponding equations
\begin{align}\nonumber
\rho&=\bigg[512 \gamma ^2 \eta  r^6 \varphi ^4 \chi ^4 \big(r^2 \varphi +1\big)^4
\big(2 \zeta  r^2 \big(3 r^2 \varphi +2\big)-3 q^2 \big(r^2 \varphi +1\big)\big) +32
\gamma r^4 \varphi ^2
\\\nonumber
&\times\chi ^3\big(r^2 \varphi +1\big)^2 \big(\zeta  r^2 \big(\eta \big(\varphi
\big(-64 \gamma \varphi +11 r^6 \varphi ^2+r^4 \varphi (96 \gamma \varphi +29)+r^2
(25
\\\nonumber
&-16 \gamma  \varphi(5 \varphi +4))\big)+7\big)-48 \gamma \varphi ^2 \big(r^2 \varphi
+1\big)\big)-2 q^2 \big(r^2 \varphi +1\big) \big(3 \eta +\varphi ^2 \big(3 \eta  r^4
\\\nonumber
&-4 \gamma  (7 \eta +6)\big)+6 \eta  r^2 \varphi \big)\big)+\chi \big((7 \eta +6) q^2
\big(r^2 \varphi +1\big)^3+\zeta  r^2 \big(-6 r^2 \varphi \big(\varphi \big(-32
\gamma
\\\nonumber
&+r^4 \varphi +r^2 (96 \gamma \varphi +3)\big)+3\big)+\eta \big(r^2 \big(\varphi
\big(208 \gamma \varphi +14 r^6 \varphi ^2+r^4 \varphi (22-9 \varphi )+r^2
\\\nonumber
&\times(2-\varphi(304 \gamma \varphi +25))-23\big)-6\big)-7\big)-6\big)\big) +2 r^2
\chi ^2 \big(\zeta  r^2 \big(\eta \big(\varphi  \big(-120 \gamma  \varphi
\\\nonumber
&+5 r^{10} \varphi ^4+r^8 \varphi ^3 (208 \gamma \varphi +23)+2 r^6 \varphi ^2 (21-76
\gamma (\varphi -2) \varphi )-2 r^4 \varphi  (4 \gamma  \varphi(\varphi  (96
\\\nonumber
&\times \gamma \varphi +53)+2)-19)+r^2 (8 \gamma \varphi (\varphi  (128 \gamma
\varphi -49)-14)+17)\big)+3\big)-96 \gamma \varphi ^2
\\\nonumber
&\times\big(r^2 \varphi  \big(\varphi \big(-8 \gamma+r^4 \varphi +3 r^2 (8 \gamma
\varphi +1)\big)+3\big)+1\big)\big)-q^2 \big(r^2 \varphi +1\big)^3 \big(3 \eta
+\varphi ^2
\\\nonumber
&\times\big(3 \eta  r^4-16 \gamma  (7 \eta +6)\big)+6 \eta  r^2 \varphi
\big)\big)+\zeta r^2 \big(7 \eta -(11 \eta +18) r^2 \varphi
+6\big)\bigg]\bigg[r^4\chi \big(r^2
\\\label{3b}
&\times \varphi +1\big)^3 \big(16 \gamma  r^2 \varphi ^2 \chi +1\big)^2 \big(\eta
\big(-8 \eta +2 (4 \eta +3) \chi  \big(r^3 \varphi
+r\big)^2-15\big)-6\big)\bigg]^{-1},
\\\nonumber
p_{r}&=\frac{1}{\eta  (\eta +1)}2 \eta \bigg[ \big(\frac{q^2}{r^4}+\big\{2 \zeta
\varphi \big(16 \gamma \varphi \chi  \big(2 r^2 \varphi  \big(4 \gamma  \varphi \chi
\big(3 r^2 \varphi -1\big)+1\big)-1\big)+1\big)\big\}
\\\nonumber
&\times\big\{\chi \big(r^2 \varphi +1\big)^3 \big(16 \gamma r^2 \varphi ^2 \chi
+1\big)^2\big\}^{-1}\big)+\big\{2 (4 \eta +3) \big(\chi \big(r^2 \varphi +1\big)^3
\big(q^2-\zeta  r^2\big)
\\\nonumber
&-3 \zeta  r^4 \varphi +\zeta r^2\big)\big\}\big\{r^4 \chi \big(r^2 \varphi
+1\big)^3\big\}^{-1}-\big\{(\eta +1) (11 \eta +6) \big(-\frac{2 (5 \eta +3) q^2}{r^4}
\\\nonumber
&+\frac{1}{r^4}\big\{\eta  \big(3 q^2 \big(2 \chi \big(r^3 \varphi
+r\big)^2+1\big)+\big\{\zeta  r^2 \big(r^2 \big(-64 \gamma  r^2 \varphi ^2 \chi ^3
\big(3 r^2 \varphi +2\big)
\\\nonumber
&\times\big(r^2 \varphi +1\big)^4-2 \big(r^2 \varphi \big(\varphi \big(-16 \gamma
(\varphi +4)+5 r^4 \varphi +r^2 (96 \gamma \varphi +13)\big)+11\big)+3\big)
\\\nonumber
&\times \big(r^2 \varphi \chi +\chi \big)^2-\varphi \chi \big(32 \gamma \varphi +14
r^6 \varphi ^2-r^4 (\varphi -22) \varphi +r^2 (\varphi (192 \gamma \varphi -1)+2)
\\\nonumber
&+1\big) -9 \varphi +6 \chi \big)-\chi +1\big)\big\}\big\{\chi \big(r^2 \varphi
+1\big)^3 \big(16 \gamma r^2 \varphi ^2 \chi +1\big)\big\}^{-1}\big)\big\} +2 \zeta
\\\nonumber
&\times \big(-\big\{16\gamma  \eta \varphi ^2 \big(r^2 \varphi
-1\big)\big\}\big\{\big(r^2 \varphi +1\big)^3 \big(16 \gamma r^2 \varphi ^2 \chi
+1\big)\big\}^{-1}+\{16 \gamma \eta \varphi ^2\}
\\\nonumber
&\times\{\big(r^2 \varphi +1\big)^2 \big(16 \gamma  r^2 \varphi ^2 \chi
+1\big)^2\}^{-1}+\{-4 \eta +(10 \eta +9) r^2 \varphi -3\}\{r^2 \chi
\\\label{3c}
&\times\big(r^2 \varphi +1\big)^3\}^{-1}+\frac{4 \eta
+3}{r^2}\big)\big)\big\}\big\{\eta \big(-8 \eta +(4 \eta +3) \chi \big(r^3 \varphi
+r\big)^2\big)\big\}^{-1}\bigg],
\\\nonumber
p_{t}&=\bigg[\chi  \big(q^2 \big(9 \eta ^2 \big(r^2 \varphi +1\big)^3+17 \eta
\big(r^2 \varphi +1\big)^3+6 r^2 \varphi  \big(r^2 \varphi \big(r^2 \varphi
+3\big)+3\big)\big)+\zeta  r^2
\\\nonumber
&\times\big(-r^2 \varphi ^2 \big(16 \gamma(\eta +1) (13 \eta +12)+\eta  r^2
\big(-\eta +2 (7 \eta +8) r^2+2\big)\big)-\eta r^2 \varphi \big(\eta
\\\nonumber
&+2 (5 \eta +4) r^2+4\big)-\eta \big(\eta +2 \eta r^2+2\big)+r^4 \varphi ^3 \big(16
\gamma (\eta (3 \eta +25)+24)+\eta  r^2 \big(\eta
\\\nonumber
&-2 (3 \eta +4) r^2\big)\big)\big)\big)-32 \gamma r^4 \varphi ^2 \chi ^3\big(r^2
\varphi +1\big)^2 \big(2 q^2 \big(r^2 \varphi +1\big) \big(\eta (5 \eta +3)+\varphi
^2
\\\nonumber
&\times\big(\eta (5 \eta +3) r^4-4 \gamma  (\eta (9 \eta +17)+6)\big)+2 \eta (5 \eta
+3) r^2 \varphi \big)+\zeta \eta  r^2 \big(-\eta +(\eta +7)
\\\nonumber
&\times r^2 \varphi+r^2 \varphi ^3 \big(-8 (2 \gamma \eta +\gamma )+(3 \eta +5)
r^4+96 \gamma (\eta +1) r^2\big)+\varphi ^2 \big(8 \gamma +(5 \eta +11) r^4
\\\nonumber
&-64 \gamma (\eta +1) r^2\big)+1\big)\big)-2 r^2 \chi ^2 \big(q^2 \big(r^2 \varphi
+1\big)^3 \big(\eta (5 \eta +3)+\varphi ^2 \big(\eta (5 \eta +3) r^4-16
\\\nonumber
&\times \gamma  (\eta(9 \eta +17)+6)\big)+2 \eta  (5 \eta +3) r^2 \varphi \big)+\zeta
r^2 \big(-\eta ^2+\eta (2-3 \eta ) r^2 \varphi +r^2 \varphi ^4
\\\nonumber
&\times\big(\gamma ^2 (\eta +1) (4 \eta +3)+\eta (3 \eta +8) r^6+8 \gamma \eta r^2
\big(-5 \eta +2 (3 \eta +7) r^2+1\big)\big)+2 \eta r^2 \varphi ^3
\\\nonumber
&\times\big(-4 \gamma  (\eta -5)+(\eta +6) r^4-8 \gamma  (9 \eta +7) r^2\big)-2 \eta
\varphi ^2\big(-4 \gamma (\eta +3)+(\eta -4) r^4
\\\nonumber
&+56 \gamma  (\eta +1) r^2\big)+r^4 \varphi ^5 \big(-768 \gamma ^2 (\eta +1) (\eta
+3)+\eta  (\eta +2) r^6+8 \gamma  \eta r^2 \big(-3 \eta
\\\nonumber
&+2 (5 \eta +7) r^2-1\big)\big)\big)\big)-512 \gamma ^2 \eta r^6 \varphi ^4 \chi ^4
\big(r^2 \varphi +1\big)^4 \big((5 \eta +3) q^2 \big(r^2 \varphi +1\big)+\zeta
\\\nonumber
&\times(2 \eta +3) r^4 \varphi +\zeta  r^2\big)+6 q^2 \chi +\zeta r^2 \big(\eta
(\eta +2)+(\eta (3 \eta +14)+12) r^2 \varphi \big)\bigg]\bigg[r^4
\\\nonumber
&\times\chi \big(r^2 \varphi +1\big)^3 \big(16 \gamma r^2 \varphi ^2 \chi +1\big)^2
\big(\eta \big(-8 \eta ^2-23 \eta +2 (\eta +1) (4 \eta +3) \chi \big(r^3 \varphi
+r\big)^2
\\\label{3d}
&-21\big)-6\big)\bigg]^{-1}.
\end{align}

\subsection{Matching Conditions}

A central component of this research involves identifying the values
of the constants. This is generally achieved by matching the
interior geometry with the exterior spacetime. In our investigation,
we have specifically matched the interior spacetime to the Bardeen
black hole, as illustrated by the following expression \cite{7b}
\begin{equation}\label{9a}
ds_+^{2}= -A(r)dt^{2}+A(r)^{-1}dr^{2}+ r^{2}(d\theta^{2}+\sin^{2}\theta d\phi^{2}),
\end{equation}
where
\begin{equation}
A(r) = 1 - \frac{2 M r^2}{\left(q^2 + r^2\right)^{3/2}}.
\end{equation}
In this framework, $M$ signifies the standard mass associated with a
celestial body. When coupled with nonlinear electrodynamics, the
Einstein field equations admit the Bardeen black hole as a magnetic
solution. This is selected for the outer layer due to its ability to
prevent singularities, satisfying the conditions of flatness and
weak energy condition and maintains a regular center. Additionally,
the geometry defined by Eq.\eqref{9a} demonstrates asymptotic
behavior that aligns with
\begin{equation}\label{1}
A(r) = 1 - \frac{2M}{r} + \frac{3M q^2}{r^3} + \mathcal{O}\left(\frac{1}{r^5}\right).
\end{equation}
In this equation for $A(r)$, the presence of the $\frac{1}{r^3}$
term highlights the distinction between the Bardeen geometry and the
standard Reissner-Nordstr$\ddot{o}$m spacetime configuration. The
higher-order term, $\mathcal{O}\left(\frac{1}{r^5}\right)$, along
with its associated contributions, is disregarded due to its
negligible value. For our analysis, we approximate $A(r)$ as
\begin{equation}\nonumber
A(r) \approx 1 - \frac{2M}{r} + \frac{3Mq^2}{r^3}.
\end{equation}

To investigate the physical properties, it is essential to align the
interior and exterior geometries of the stars at the surface $r =
R$. Ensuring the continuity of the interior and exterior spacetimes
at the boundary requires $g_{tt}$, $g_{rr}$ and
$\frac{\partial}{\partial r}(g_{tt})$ to remain continuous. This
condition results in the following set of equations. Through the
preservation of continuity in the metric coefficients outlined in
Eqs.\eqref{1a} and \eqref{9a}, we conclude
\begin{eqnarray}\label{31} \frac{3 M q^2}{R^3}-\frac{2 M}{R}+1=\chi  \big(R^2
\varphi +1\big)^2,\\\label{32} \bigg(\frac{3 M q^2}{R^3}-\frac{2
M}{R}+1\bigg)^{-1}=16 \gamma R^2 \varphi ^2 \chi +1.
\end{eqnarray}
Differentiating Eq.\eqref{31} with respect to $R$, we obtain
\begin{eqnarray}\label{33}
\frac{2 M}{R^2}-\frac{9 M q^2}{R^4}=4 R \varphi  \chi  \big(R^2 \varphi +1\big).
\end{eqnarray}
Solving these equations simultaneously, the expressions for $\chi$,
$\gamma$ and $\varphi$ are derived as
\begin{eqnarray}\label{34}
\chi &=&\frac{\big(21 M q^2-10 M R^2+4 R^3\big)^2}{16 R^3 \big(3 M q^2-2 M
R^2+R^3\big)},
\\\label{35} \gamma &=&\frac{2 R^7-3 q^2 R^5}{M \big(9 q^2-2 R^2\big)^2},
\\\label{36} \varphi&=&\frac{M \big(2 R^2-9 q^2\big)}{21 M q^2 R^2-10 M R^4+4 R^5}.
\end{eqnarray}

\section{Physical Features}

In this section, we assess the viability of charged compact stars by
investigating the fundamental physical properties. Graphical
representations are employed to illustrate the applicability and
effectiveness of the proposed model for such astrophysical systems.
The analysis encompasses key physical parameters, including the
$\rho$, $p_r$, $p_t$, anisotropy and the gradients of $\rho$, $p_r$
and $p_t$. Additionally, we examine energy conditions, the EoS
parameter, the Tolman-Oppenheimer-Volkoff (TOV) equation,
mass-radius relation, compactness and redshift. Stability is
evaluated using criteria such as causality conditions and the
adiabatic index. For graphical analysis, various values of $\zeta =
1.9, 1.7, 1.5, 1.3, 1.1$ are considered, along with $\kappa = 8\pi$.

\subsection{Dynamics of Fluid Parameters}

Analyzing fluid properties like $\rho$ and pressure is crucial for
comprehending the internal structure and dynamics of astrophysical
objects. These parameters peak at the core, where the immense
density counterbalances gravitational forces, preserving the star
structural integrity. Significant deviations or negative values can
disrupt this balance, possibly leading to structural instability.
Investigating how these parameters evolve across various theoretical
frameworks not only deepens our comprehension of astrophysical
phenomena but also enhances observational techniques and the
analysis of data gathered from telescopes and other instruments.
Figure \textbf{1} illustrates that matter density is the highest at
the core and gradually diminishes with increasing radial distance,
underscoring the compact and dense nature of charged stars in the
Bardeen framework. Likewise, $p_{r}$ steadily decreases outward from
the core and approaches to zero at the boundary, ensuring overall
stability. In Figure \textbf{2}, the plots illustrate the densely
packed configuration of charged compact stars within the Bardeen
framework. The $p_{t}$ starts at zero at the core but becomes
negative at greater radial distances, indicating anisotropic
behavior in the outer layers. These results emphasize the importance
of fluid parameters in analyzing the dense and stable
characteristics of charged compact stars under Bardeen conditions.
\begin{figure}\center
\epsfig{file=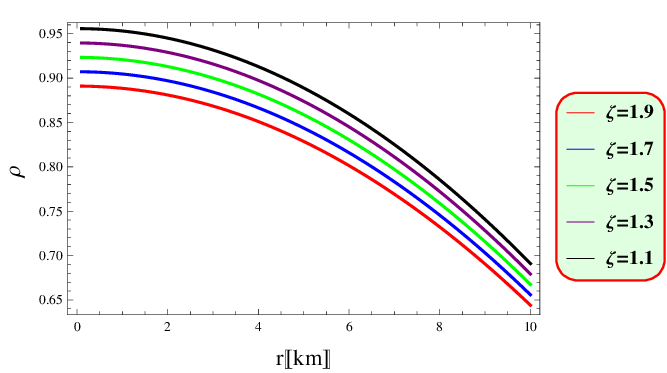,width=.47\linewidth} \epsfig{file=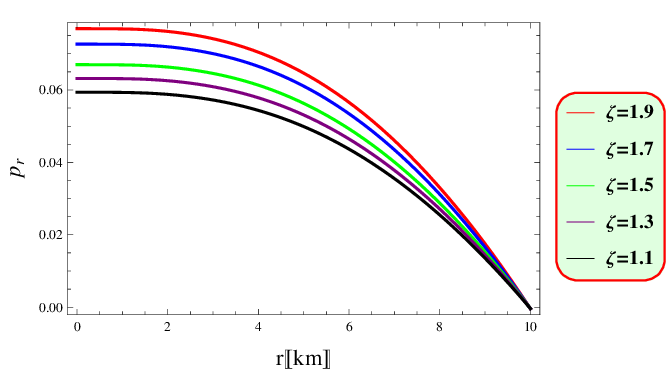,width=.47\linewidth}
\epsfig{file=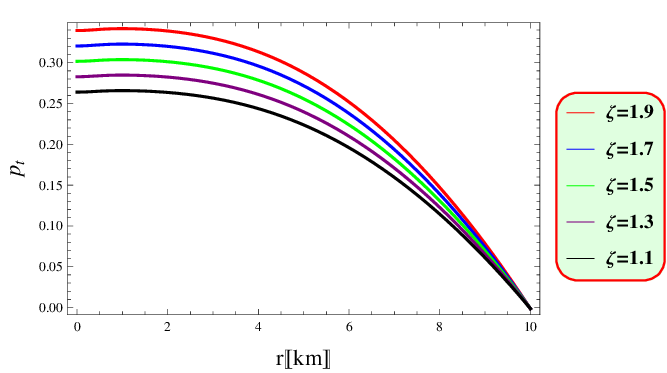,width=.47\linewidth}\caption{Graphs of $\rho$, $p_r$ and $p_t$ as
functions of $r$.}
\end{figure}
\begin{figure}\center
\epsfig{file=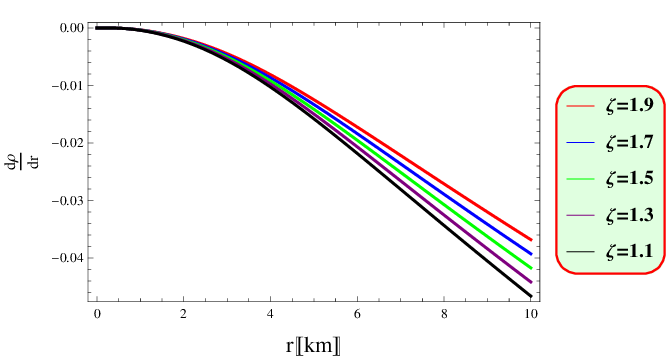,width=.47\linewidth} \epsfig{file=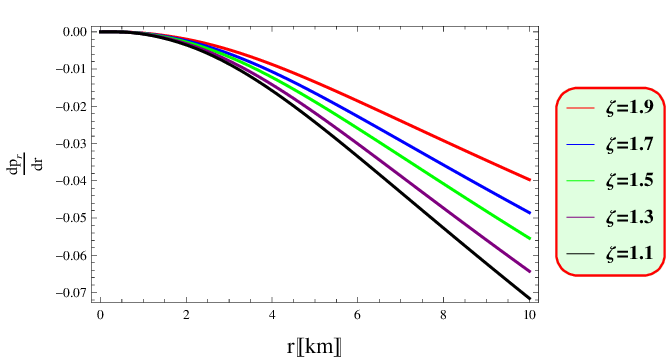,width=.47\linewidth}
\epsfig{file=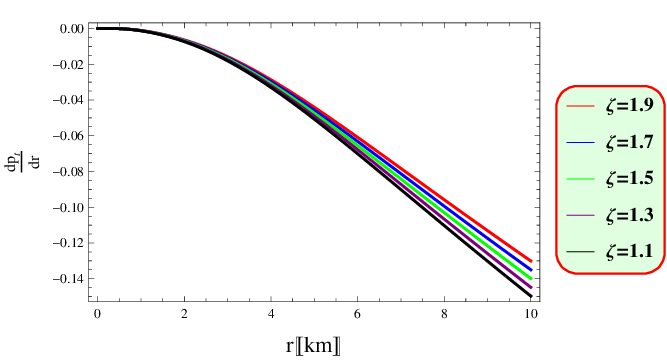,width=.47\linewidth}\caption{Graphs of $\frac{d\rho}{dr}$,
$\frac{dp_r}{dr}$ and $\frac{dp_t}{dr}$ with respect to $r$.}
\end{figure}

\subsection{ Anisotropy}

Anisotropy describes the condition in which the physical properties
of a material, structure or space vary depending on the direction of
observation or measurement. In astrophysics, this concept is
particularly significant in the study of stellar objects, where
internal dynamics, matter distribution, and high-density effects
result in varying pressures along radial and tangential directions.
The anisotropic function, defined as $\Delta = p_t - p_r$, is a key
tool for analyzing these variations: when $\Delta = 0$, the pressure
is uniformly distributed, indicating isotropy. If $\Delta > 0$, it
corresponds to an outward directed anisotropic force, whereas
$\Delta < 0$ indicates an inward directed force. This characteristic
is critical in understanding the stability and behavior of compact
stars. As depicted in Figure \textbf{3}, the presence of charge
induces a repulsive force, represented by a positive value that
gradually decreases. This behavior is essential for supporting
large-scale structures and mitigating the risk of gravitational
collapse.
\begin{figure}\center
\epsfig{file=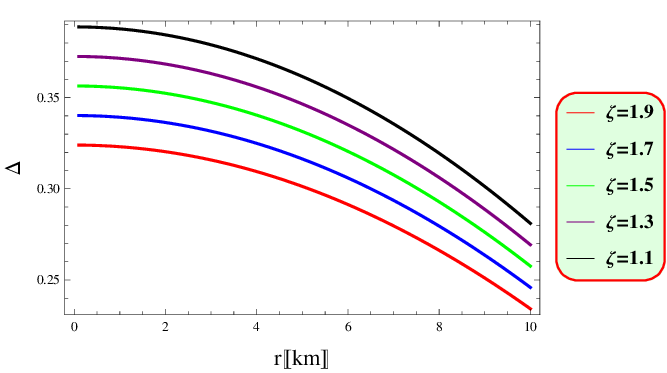,width=.47\linewidth} \caption{Graph of $\Delta$ against $r$.}
\end{figure}

\subsection{Energy conditions}

Energy conditions are essential for comprehending the physical
properties and dynamics of compact objects in cosmology. They
provide critical constraints that assess the feasibility and
reliability of stellar models. In the context of charged compact
stars, adherence to these conditions is necessary to ensure model
stability and physical validity. These are given as
\begin{enumerate}
\item The null energy condition states that:
$p_r+\rho\geq 0$ and $p_t+\rho + \frac{q^2}{4\pi r^2} \geq 0$.
\item The dominant energy condition requires that:
$\rho - p_r \geq 0$ and $\rho - p_t + \frac{q^2}{4\pi r^2} \geq 0$.
\item The weak energy condition is defined as:
$\rho + \frac{q^2}{4\pi r^2} \geq 0$, $p_r+\rho\geq 0$ and $p_t+\rho +
\frac{q^2}{4\pi r^2} \geq 0$.
\item The strong energy condition specifies that:
$p_r+\rho\geq 0$ and $p_t+\rho + \frac{q^2}{4\pi r^2} \geq 0$, $p_r+\rho + 2p_t +
\frac{q^2}{4\pi r^2} \geq 0$.
\end{enumerate}
These conditions help in distinguishing ordinary matter from exotic
matter. When all conditions are fulfilled, the matter is regarded as
ordinary, whereas failing to meet them classifies the matter as
exotic. These conditions collectively play a vital role in analyzing
the stability and dynamics of cosmological models and compact
objects. The results, as shown in Figure \textbf{4}, indicate that
all the energy conditions are satisfied. This demonstrates that the
anisotropic charged compact star described in the Bardeen framework
represents a physically plausible model with a regular and
non-exotic matter composition.
\begin{figure}\center
\epsfig{file=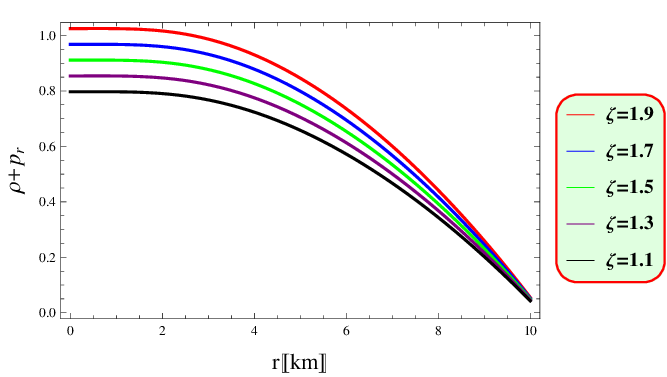,width=.47\linewidth}\epsfig{file=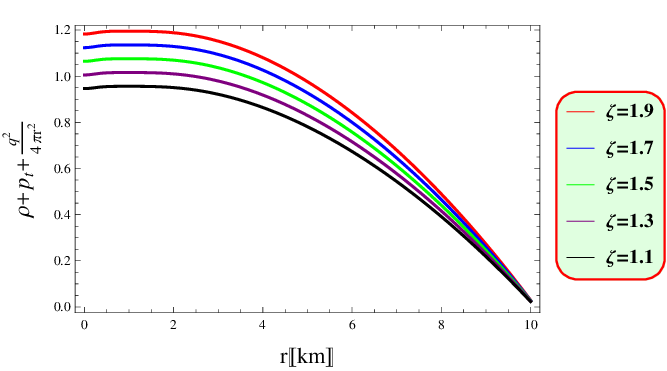,width=.47\linewidth}
\epsfig{file=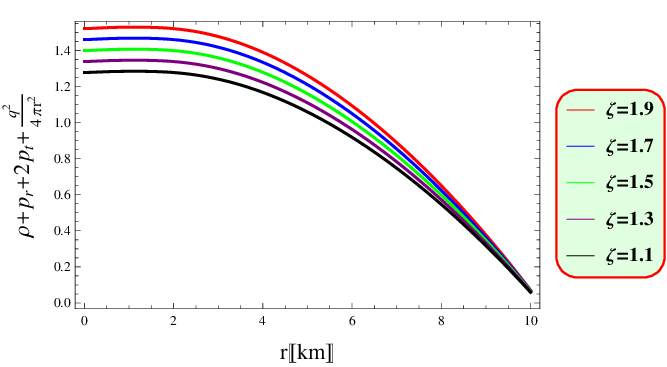,width=.47\linewidth}\epsfig{file=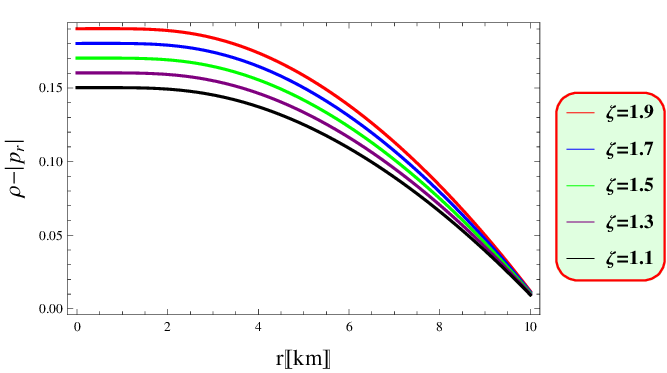,width=.47\linewidth}
\epsfig{file=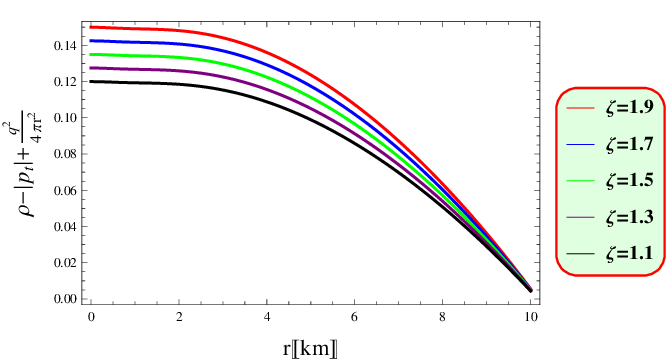,width=.47\linewidth} \caption{Graphs of energy conditions as
functions of $r$.}
\end{figure}

\subsection{Equation of State Parameter}

The EoS parameters, $\omega_r = \frac{p_r}{\rho}$ and $\omega_t =
\frac{p_t}{\rho}$, are essential in defining the relationship
between pressure and energy density within compact stars. These
parameters provide valuable insights into the star matter
distribution and thermodynamic behavior. Ensuring the physical
consistency of the stellar model requires that the EoS parameters
fall between 0 and 1. Figure \textbf{5} illustrates an upward trend
in both $\omega_r$ and $\omega_t$ while confirming that their values
remain within the expected range, validating the physical viability
and stability of the configuration in this analysis.
\begin{figure}\center
\epsfig{file=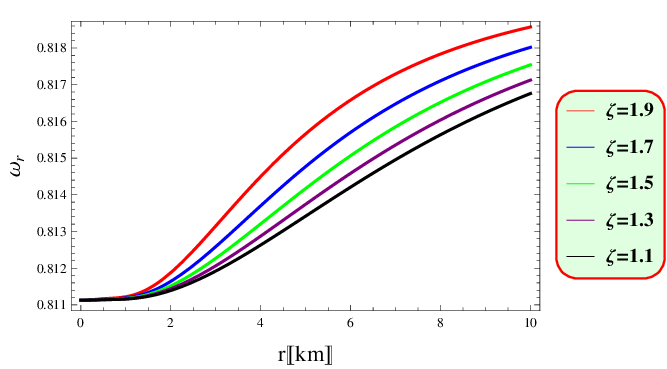,width=.47\linewidth} \epsfig{file=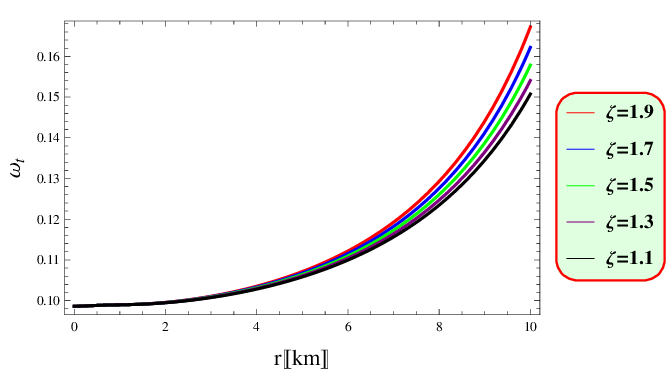,width=.47\linewidth}
\caption{Graphs of EoS against $r$.}
\end{figure}

\subsection{Equilibrium condition}

This section delves into the equilibrium state of a charged compact
object within the Bardeen framework. Equilibrium occurs when all the
forces acting on a system are in perfect opposition, resulting in a
net force of zero. To analyze this balance, the TOV equation is
utilized. This equation is a fundamental tool for exploring how
gravitational collapse is counteracted by internal pressure,
ensuring stability \cite{7d}. It also offers valuable insights into
the internal dynamics and structural properties of dense stars. By
incorporating the effects of charge, the TOV equation can be
formulated as \cite{7e}
\begin{equation*}
\frac{d p_r}{d r} + \frac{\alpha^{\prime}(r)}{2} (p_r+\rho) + \frac{2}{r} (p_t - p_r)
+ \frac{\sigma(r)\mathbb{E}(r)e^{\beta(r)}}{2} = 0.
\end{equation*}
This can be rewritten in terms of force components as
\begin{equation*}
F_a + F_g + F_e + F_h = 0,
\end{equation*}
where each term represents a distinct force
\begin{itemize}
\item Anisotropic force: $F_a = \frac{2}{r} (p_t - p_r),$
\item Gravitational force: $F_g = \frac{\alpha^{\prime}(r)}{2} (\rho+p_r),$
\item Electric force: $F_e = \frac{\sigma(r)\mathbb{E}(r)e^{\beta(r)}}{2},$
\item Hydrostatic force: $F_h = \frac{d p_r}{d r}$.
\end{itemize}
Figure \textbf{6} graphically illustrates the behavior of these
forces. Positive values characterize the anisotropic and
gravitational forces, while the electric and hydrostatic forces are
negative. These opposing contributions collectively neutralize each
other, confirming that the model achieves a state of equilibrium
\cite{7f}.
\begin{figure}\center
\epsfig{file=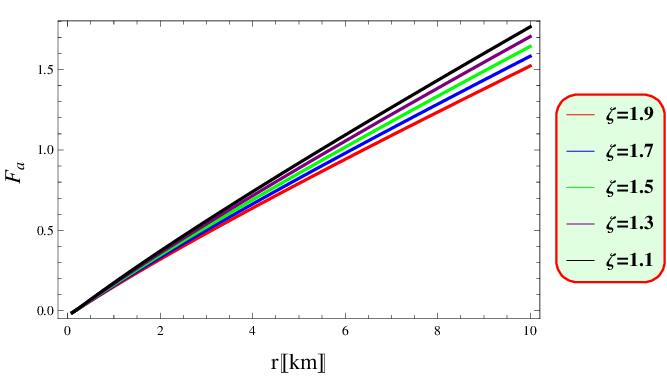,width=.47\linewidth}\epsfig{file=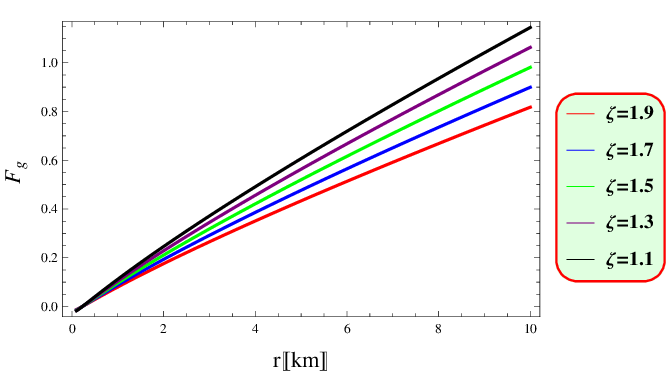,width=.47\linewidth}
\epsfig{file=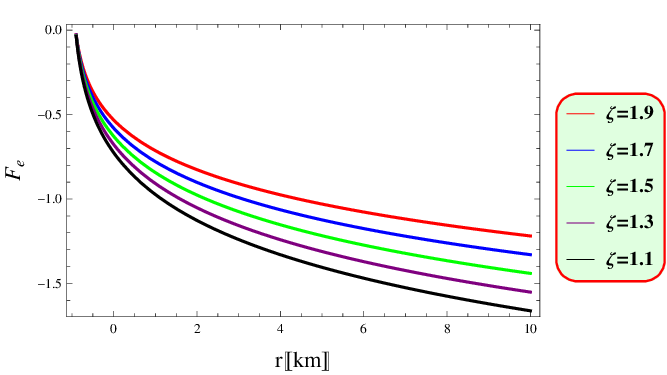,width=.47\linewidth}\epsfig{file=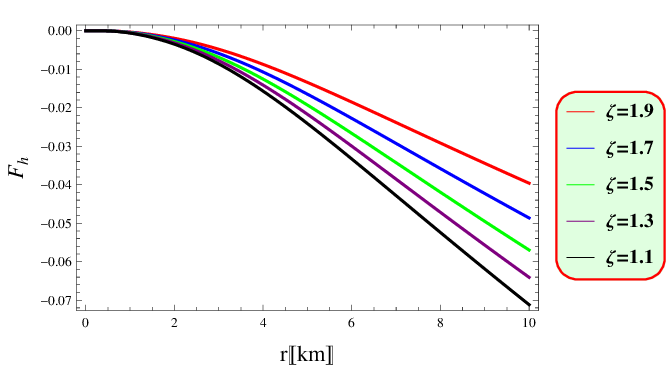,width=.47\linewidth}
\caption{Plots of Forces against $r$.}
\end{figure}

\subsection{Mass-Radius, Compactness and Redshift}

The condition $\frac{2M}{R} \leq \frac{8}{9}$ must be met by the
mass to radius ratio of a compact star. Here, $R$ represents the
radius in Schwarzschild coordinates and $M$ corresponds to the fluid
total mass, calculated at the point where the pressure becomes zero
\cite{7c}. The mass function is given as
\begin{equation}\label{37}
M(r) = 4\pi \int_{0}^{r} \mathbf{r}^2 \rho \, d\mathbf{r}.
\end{equation}
As $r \to 0$, the behavior of the mass function reveals that $M(r)
\to 0$, indicating it remains regular and well-defined at the
center, even under the Bardeen framework. Additionally, as
illustrated in Figure \textbf{7}, the mass function exhibits a
steady and monotonic increase with $r$.
\begin{figure}\center
\epsfig{file=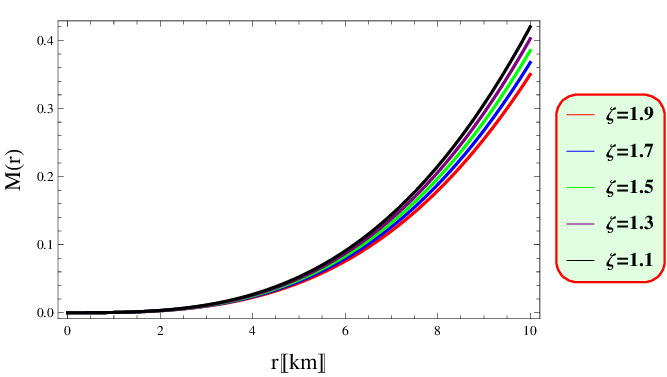,width=.47\linewidth} \caption{Plot of $M(r)$ against $r$.}
\end{figure}

In astrophysics, the concept of compactness refers to the density of
matter within an astronomical body, defined mathematically as $U(r)
= \frac{M(r)}{r}$. Compact objects are characterized by high
compactness, where significant mass is concentrated within a limited
volume. To ensure stability, the compactness must not exceed the
critical threshold of $\frac{4}{9}$ \cite{7c}. If this limit is
surpassed, gravitational collapse may occur, potentially leading to
the formation of event horizons. Figure \textbf{8} depicts a gradual
increase in compactness, remaining well within the stability limits.
\begin{figure}\center
\epsfig{file=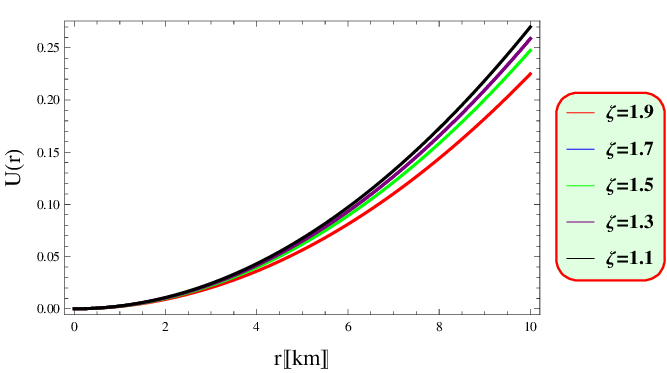,width=.47\linewidth} \caption{Plot of $U(r)$ against $r$.}
\end{figure}

Compact objects are known for their significant gravitational
redshift, a result of their strong gravitational fields. As
radiation or light travels away from these dense objects, it
experiences a loss of energy, causing its wavelength to stretch.
This phenomenon, called surface redshift, provides critical insights
into the properties of the emitted light and the gravitational
intensity at the object surface. Consequently, surface redshift acts
as an important parameter in the evaluation of dense stellar
remnants. Its mathematical representation is given as
\begin{equation}\nonumber
Z(r) = \big(1 - 2U(r)\big)^{-\frac{1}{2}} - 1.
\end{equation}
In the case of an anisotropic configuration, the surface redshift
must remain below $Z(r) \leq 5.211$ for the compact star to be
deemed physically viable \cite{7g}. As depicted in Figure
\textbf{9}, the redshift function adheres to this requirement,
staying positive and finite. It shows an increasing trend towards
the surface, providing important information about the stability and
feasibility of the model.
\begin{figure}\center
\epsfig{file=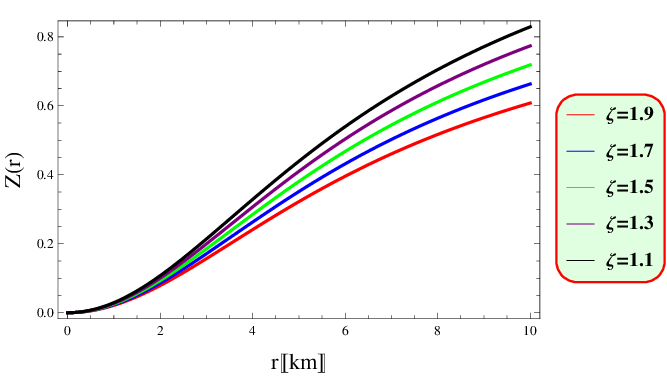,width=.47\linewidth} \caption{Graph of $Z(r)$ against $r$.}
\end{figure}

\subsection{Stability}

Understanding the physical features and stability of celestial
entities is a key aspect of gravitational physics. Stability is
crucial to ensure the structural integrity and consistency of cosmic
bodies. To assess this, the causality constraint is employed, which
dictates that no signal can travel faster than the speed of light.
For a compact star to remain stable, the radial and tangential sound
speeds, represented as $v_r^2 = \frac{d{p}_{r}}{d\rho}$ and $v_t^2 =
\frac{d{p}_{t}}{d\rho}$, must fall within the interval $[0, 1]$
\cite{7h}. These constraints on sound speeds are essential for the
star equilibrium. Figure \textbf{10} demonstrates that the
anisotropic charged compact star, developed within the Bardeen
framework, fulfills the necessary criteria. This validates that MGT
supports the formation of a stable and physically viable compact
star structure.
\begin{figure}\center
\epsfig{file=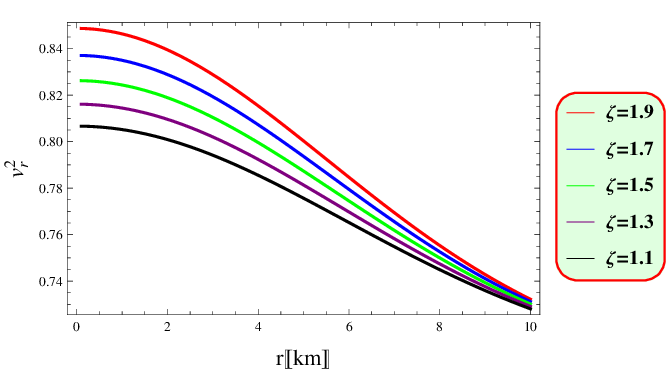,width=.47\linewidth} \epsfig{file=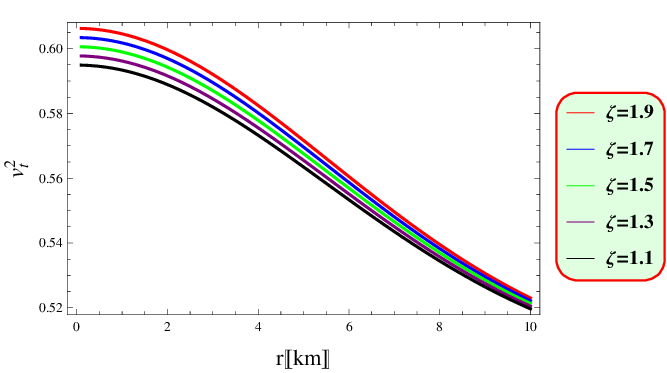,width=.47\linewidth}
\caption{Plots of causality conditions against $r$.}
\end{figure}

To assess the stability of the solutions, the cracking method is
applied, a technique introduced by Herrera \cite{8a}, which relies
on the condition $0\leq\mid v_t^2-v_r^2\mid\leq 1$. When this
condition is met, it signifies that cosmic structures are stable and
can preserve their configuration over time. In contrast, a violation
of this criterion indicates instability, which could lead to
structural collapse. Figure \textbf{11} demonstrates that the
cracking condition remains within the interval $[0, 1]$, confirming
the stability of the anisotropic charged compact star described in
the Bardeen model.
\begin{figure}\center
\epsfig{file=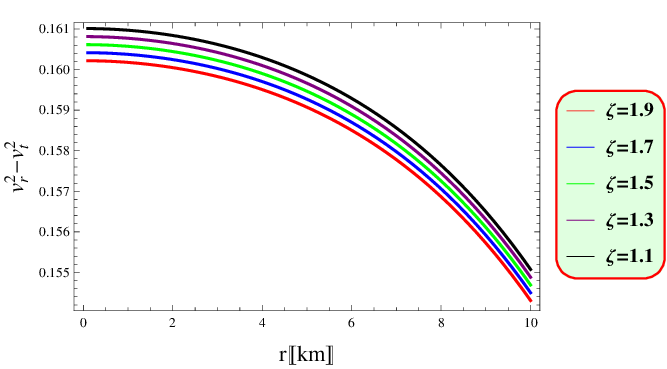,width=.47\linewidth} \caption{Plot of Herrera cracking against
$r$.}
\end{figure}

As a fundamental parameter, the adiabatic index $\Gamma$ plays a
vital role in determining the stability of dense stellar remnants
and offers important information regarding the fluid dynamics within
these systems. It is divided into two components: radial $\Gamma_r$
and tangential $\Gamma_t$, formally defined as
\begin{equation}
\Gamma_r = \frac{\nu_r^2 (\rho+p_r)}{p_r}, \quad \Gamma_t = \frac{\nu_t^2 ( \rho+p_t
)}{p_t}.
\end{equation}
For compact astrophysical objects to remain stable and resist
gravitational collapse, both $\Gamma_r$ and $\Gamma_t$ must exceed
$\frac{4}{3}$ \cite{8b}. This index is a critical tool for examining
the structural integrity of stars and understanding the behavior of
matter in extreme conditions.
\begin{figure}\center
\epsfig{file=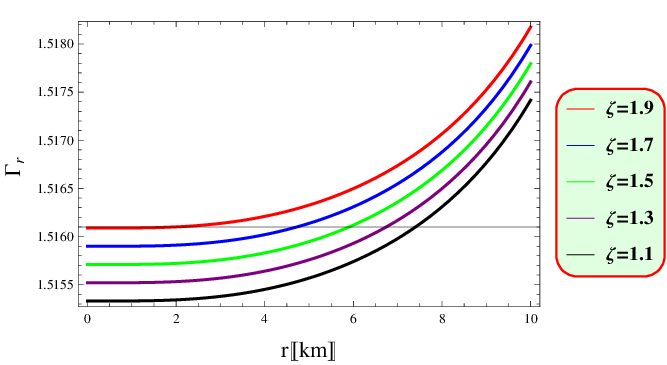,width=.47\linewidth} \epsfig{file=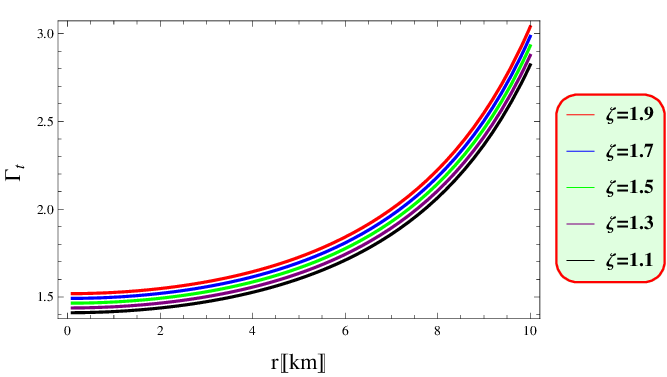,width=.47\linewidth}
\caption{Plots of $\Gamma_r$ and $\Gamma_t$ against $r$.}
\end{figure}
Figure \textbf{12} demonstrates that the system remains stable by
meeting the required conditions. Such stability is vital for the
formation of compact stars, enabling them to resist gravitational
collapse and preserve energy with minimal dissipation.

\section{Concluding Remarks}

Modified theories are pivotal in astrophysics for understanding the complexities of
stellar configurations. Among these, $f(\mathfrak{Q}, \mathcal{T})$ gravity has
attracted considerable interest due to its unique integration of non-metricity and
matter, producing notable results in thermodynamics, cosmology and the study of
relativistic stars. This theory has shown significant promise in astrophysical
research. Our study focuses on exploring the characteristics of charged stellar
structures in this MGT. The field equations are solved by employing a linear model
$f(\mathfrak{Q}, \mathcal{T}) = \zeta \mathfrak{Q} + \eta \mathcal{T}$ and the
Finch-Skea metric to simplify the analysis. We use Bardeen geometry for external
spacetime for matching conditions that allow a comparison of internal and external
geometries, facilitating the determination of unknown constants critical in
understanding the properties of charged compact stars. Stability and viability are
verified through graphical analysis for specific $\zeta$ values. The main findings of
this study are outlined as follows.
\begin{itemize}
\item
We have observed that the fluid variables $\rho$, $p_r$ and $p_t$
are densely concentrated near the core of the charged compact star,
highlighting a stable central region. This specific distribution of
fluid properties plays a crucial role in maintaining the overall
structural stability of the star. Additionally, the gradual decrease
of these variables towards the outer boundary supports the stability
and feasibility of the anisotropic charged star. That $p_{r}$ drops
to zero at the surface boundary further supports the physical
consistency of the stellar model (Figure \textbf{1}). The negative
gradient in matter density confirms a dense and compact stellar
configuration (Figure \textbf{2}).
\item
Anisotropic pressure is observed to be positive and gradually
diminishes outward, contributing significantly to the stability and
formation of charged compact stellar objects. This outward force
plays a key role in maintaining the structural integrity of such
stars (Figure \textbf{3}).
\item
The EoS parameter stays confined within the interval of 0 to 1,
validating the model physical coherence and reliability (Figure
\textbf{4}).
\item
All energy conditions are met for different values of $\zeta$,
indicating the presence of ordinary matter. Consequently, the model
is confirmed to be physically viable (Figure \textbf{5}).
\item
The mass function shows a progressive increase with the radial
coordinate, reflecting a realistic and physically consistent
distribution of mass for the charged compact star (Figure
\textbf{6}).
\item
It is observed that $F_a$ and $F_g$ have positive values, whereas
$F_e$ and $F_h$ are negative. The equilibrium condition is achieved
within the charged framework as the total sum of these forces equals
zero (Figure \textbf{7}).
\item
The stability of charged compact star is checked using factors such
as compactness, redshift, causality limits, the cracking method and
the adiabatic index. The outcomes confirm that the stability
conditions are met, demonstrating the existence of a stable charged
compact star (Figures \textbf{8-12}).
\end{itemize}

Pradhan and Sahoo \cite{9b} explored the similar work in
$f(\mathfrak{Q})$ gravity using conformal factor and found insights
into their role in stellar structures. In this framework, the
stability of results is model dependent: while model-1 achieves
stability, model-2 which is linear, encounters core singularities.
Physical parameters like pressure and density exhibit central
singularity due to conformal symmetries - a significant drawback as
it fails to eliminate core singularities. In contrast, our research
demonstrates that gravitational and hydrostatic forces effectively
balance under Bardeen geometry and Finch-Skea metric, rendering our
constructed model stable and physically acceptable. In the
$f(\mathfrak{Q})$ framework, these forces fail to balance, leading
to instability in stellar equilibrium. Additionally, we have
compared our results with $f(\mathfrak{R})$ gravity \cite{9a}, a
fourth-order theory, while our second-order theory simplifies
mathematical formulation. Both approaches employ the Finch-Skea
metric and show consistent results. Our work represents a superior
theoretical framework in terms of innovation and versatility,
contributing significantly to understand the structure and behavior
of compact stellar objects.\\\\
\textbf{Data Availability Statement:} No new data were generated or
analyzed in support of this research.

\end{document}